\documentclass{article}
\usepackage{arxiv}
\usepackage[utf8]{inputenc}
\usepackage[T1]{fontenc}
\usepackage[inline]{enumitem}
\usepackage{booktabs}
\usepackage{amsmath}
\usepackage{amsfonts}
\usepackage{amssymb}
\usepackage{enumitem}
\usepackage{mathbbol}
\usepackage{mathtools}
\usepackage{hyperref}
\usepackage{url}
\usepackage{nicefrac}
\usepackage{microtype}
\usepackage{natbib}
\usepackage{multirow}

\newcommand{\eg}{\emph{e.g.}}

\title{An updated duet model for passage re-ranking}

\author{
  Bhaskar Mitra \\
  Microsoft AI \& Research \\
  Montreal, Canada \\
  \texttt{bmitra@microsoft.com} \\
   \And
 Nick Craswell \\
  Microsoft AI \& Research \\
  Redmond, USA \\
  \texttt{nickcr@microsoft.com} \\
}

\begin{document}
\maketitle

\begin{abstract}
We propose several small modifications to Duet---a deep neural ranking model---and evaluate the updated model on the MS MARCO passage ranking task. We report significant improvements from the proposed changes based on an ablation study.
\end{abstract}

\keywords{Neural information retrieval \and Passage ranking \and Ad-hoc retrieval \and Deep learning}

\section{Introduction}
\label{sec:intro}
In information retrieval (IR), traditional learning to rank \citep{Liu:2009} models estimate the relevance of a document to a query based on hand-engineered features.
The input to these models typically includes, among others, features based on patterns of exact matches of query terms in the document.
Recently proposed deep neural IR models \citep{mitra2018introduction}, in contrast, accept the raw query and document text as input.
The input text is represented as one-hot encoding of words (or sub-word components \citep{kim2016character, jozefowicz2016exploring, sennrich2015neural})---and the deep neural models focus primarily on learning latent representations of text that are effective for matching query and document.
\citet{mitra2016learning} posit that deep neural ranking models should focus on both:
\begin{enumerate*}[label=(\roman*)]
    \item representation learning for text matching, as well as on
    \item feature learning based on patterns of exact matches of query terms in the document.
\end{enumerate*}
They demonstrate that a neural ranking model called \emph{Duet}\footnote{
While \citet{mitra2016learning} propose a specific neural architecture, they refer more broadly to the family of neural architectures that operate on both term space and learned latent space as duet.
We refer to the specific architecture proposed by \citet{mitra2016learning} as Duet---to distinguish it from the general family of such architectures that we refer to as duet (note the difference in capitilization).
}---with two distinct sub-models that consider both matches in the term space (the \emph{local sub-model}) and the learned latent space (the \emph{distributed sub-model})---is more effective at estimating query-document relevance.

In this work, we evaluate a duet model on the MS MARCO passage ranking task \citep{bajaj2016ms}.
We propose several simple modifications to the original Duet architecture and demonstrate through an ablation study that incorporating these changes results in significant improvements on the passage ranking task.

\section{Passage re-ranking on MS MARCO}
\label{sec:task}
The MS MARCO passage ranking task \citep{bajaj2016ms} requires a model to rank approximately thousand passages for each query.
The queries are sampled from Bing's search logs, and then manually annotated to restrict them to questions with specific answers.
A BM25 \citep{robertson2009probabilistic} model is employed to retrieve the top thousand candidate passages for each query from the collection.
For each query, zero or more candidate passages are deemed relevant based on manual annotations.
The ranking model is evaluated on this passage re-ranking task using the mean reciprocal rank (MRR) metric \citep{craswell2009mean}.
Participants are required to submit the ranked list of passages per query for a development (dev) set and a heldout (eval) set.
The ground truth annotations for the development set are available publicly, while the corresponding annotations for the evaluation set are heldout to avoid overfitting.
A public leaderboard\footnote{\url{http://www.msmarco.org/leaders.aspx}} presents all submitted runs from different participants on this task.

\section{The updated Duet model}
\label{sec:model}
In this section, we briefly describe several modifications to the Duet model.
A public implementation of the updated Duet model using PyTorch \citep{paszke2017automatic} is available online\footnote{\url{https://github.com/dfcf93/MSMARCO/blob/master/Ranking/Baselines/Duet.ipynb}}.

\paragraph{Word embeddings}
We replace the character level $n$-graph encoding in the input of the distributed model with word embeddings.
We see significant reduction in training time given a fixed number of minibatches and a fixed minibatch size.
This change primarily helps us to train on a significantly larger amount of data under fixed training time constraints.
We initialize the word embeddings using pre-trained GloVe \citep{pennington2014glove} embeddings before training the Duet model.

\paragraph{Inverse document frequency weighting}
In contrast to some of the other datasets on which the Duet model has been previously evaluated \citep{mitra2016learning, nanni2017benchmark}, the MS MARCO dataset contains a relatively larger percentage of natural language queries and the queries are considerably longer on average.
In traditional IR models, the inverse document frequency (IDF) \citep{robertson2004understanding} of a query term provides an effective mechanism for weighting the query terms by their discriminative power.
In the original Duet model, the input to the local sub-model corresponding to a query $q$ and a document $d$ is a binary interaction matrix $X \in \mathbb{R}^{|q| \times |d|}$ defined as follows:

\begin{align}
    X_{ij} =
    \begin{cases}
        1, & \text{if } q_i = d_j\\
        0, & \text{otherwise}
    \end{cases}
\end{align}

We incorporate IDF in the Duet model by weighting the interaction matrix by the IDF of the matched terms.
We adopt the Robertson-Walker definition of IDF \citep{jones2000probabilistic} normalized to the range $[0, 1]$.

\begin{align}
    X'_{ij} =
    \begin{cases}
        \text{IDF}(q_i), & \text{if } q_i = d_j\\
        0, & \text{otherwise}
    \end{cases}
\end{align}
\begin{align}
    \text{IDF}(t) = \frac{\text{log}(N/n_t)}{\text{log}(N)}
\end{align}

Where, $N$ is the total number of passages in the collection and $n_t$ is the number of passages in which the term $t$ appears at least once.

\paragraph{Non-linear combination of local and distributed models}
\citet{zamani2018neural} show that when combining different sub-models in a neural ranking model, it is more effective if each sub-model produce a vector output that are further combined by additional multi-layer perceptrons (MLP).
In the original Duet model, the local and the distributed sub-models produce a single score that are linearly combined.
In our updated architecture, both models produce a vector that are further combined by an MLP---with two hidden layers---to generate the estimated relevance score.

\paragraph{Rectifier Linear Units (ReLU)}
We replace the Tanh non-linearities in the original Duet model with ReLU \citep{glorot2011deep} activations.

\paragraph{Bagging}
We observe some additional improvements from combining multiple Duet models---trained with different random seeds and on different random sample of the training data---using bagging \citep{breiman1996bagging}.
\section{Experiments}
\label{sec:experiment}

The MS MARCO task provides a pre-processed training dataset---called ``triples.train.full.tsv''---where each training sample consists of a triple $\langle q, p_+, p_- \rangle$, where $q$ is a query and $p_+$ and $p_-$ are a pair of passages, with $p_+$ being more relevant to $q$ than $p_-$.
Similar to the original Duet model, we employ the cross-entropy with softmax loss to learn the parameters of our model $\mathcal{M}$:

\begin{align}
    \mathcal{L} &= \mathbb{E}_{q,p_+,p_- \sim \theta} [\ell(\mathcal{M}_{q, p_+} - \mathcal{M}_{q, p_-})] \\
    \text{where,}\;\ell(\Delta) &= \text{log}(1 + e^{-\sigma\cdot\Delta}) \label{eqn:loss}
\end{align}

Where, $\mathcal{M}_{q, p}$ is the relevance score for the pair $\langle q, p \rangle$ as estimated by the model $\mathcal{M}$.
Note, that by considering a single negative passage per sample, our loss is equivalent to the RankNet loss \citep{burges2005learning}.

We use the Adam optimizer with default parameters and a learning rate of $0.001$.
We set $\sigma$ in Equation \ref{eqn:loss} to $0.1$ and dropout rate for the model to $0.5$.
We trim all queries and passages to their first $20$ and $200$ words, respectively.
We restrict our input vocabulary to the $71,486$ most frequent terms in the collection and set the size of all hidden layers to $300$.
We use minibatches of size 1024 and train the model for 1024 minibatches.
Finally, for bagging we train eight different Duet models with different random seeds and on different samples of the training data.
We train and evaluate our models using a Tesla K40 GPU---on which it takes a total of only $1.5$ hours to train each single Duet model and to evaluate it on both dev and eval sets.

\section{Results}
\label{sec:result}

\begin{table}[t]
    \centering
    \caption{Comparison of the different Duet variants and other state-of-the-art approaches from the public MS MARCO leaderboard. The update Duet model---referred to as Duet v2---benefits significantly from the modifications proposed in this paper.}
    \begin{tabular}{lcc}
    \hline
    \hline
        \multirow{2}{*}{\textbf{Model}} & \multicolumn{2}{c}{\textbf{MRR@10}} \\
        & Dev & Eval \\
        \hline
        \multicolumn{3}{l}{\textbf{Other approaches}} \\
        BM25 & $0.165$ & $0.167$ \\
        Single CKNRM \citep{dai2018convolutional} model & $0.247$ & $0.247$ \\
        Ensemble of 8 CKNRM \citep{dai2018convolutional} models & $0.290$ & $0.271$ \\
        IRNet (a proprietary deep neural model) & $0.278$ & $0.281$ \\
        BERT \citep{nogueira2019passage} & $0.365$ & $0.359$ \\
        \hline
        \multicolumn{3}{l}{\textbf{Duet variants}} \\
        Single Duet v2 w/o IDF weighting for interaction matrix & $0.163$ & - \\
        Single Duet v2 w/ Tanh non-linearity (instead of ReLU) & $0.179$ & - \\
        Single Duet v2 w/o MLP to combine local and distributed scores & $0.208$ & - \\
        Single Duet v2 model & $0.243$ & $0.245$ \\
        Ensemble of 8 Duet v2 models & $0.252$ & $0.253$ \\
        \hline
        \hline
    \end{tabular}
    \label{tbl:results}
\end{table}

Table \ref{tbl:results} presents the MRR@$10$ corresponding to all the Duet variants we evaluated on the dev set.
The updated Duet model with all the modifications described in Section \ref{sec:model}---referred hereafter as Duet v2---achieves an MRR@$10$ of $0.243$.
We perform an ablation study by leaving out one of the three modifications---\begin{enumerate*}[label=(\roman*)]
    \item IDF weighting for interaction matrix, 
    \item ReLU non-linearity instead of Tanh, and
    \item LP to combine local and distributed scores,
\end{enumerate*}---out at a time.
We observe a $33\%$ degradation in MRR by not incorporating the IDF weighting alone.
It is interesting to note that the Github implementations\footnote{
https://github.com/thunlp/Kernel-Based-Neural-Ranking-Models
} of the KNRM \citep{xiong2017end} and CKNRM \citep{dai2018convolutional} models also indicate that their MS MARCO submissions incorporated IDF term-weighting---potentially indicating the value of IDF weighting across multiple architectures.
Similarly, we also observe a $26\%$ degradation in MRR by using Tanh non-linearity instead of ReLU.
Using a linear combination of scores from the local and the distributed model instead of combining their vector outputs using an MLP results in $14\%$ degradation in MRR.
Finally, we observe a $3\%$ improvement in MRR by ensembling eight Duet v2 models using bagging.
We also submit the individual Duet v2 model and the ensemble of eight Duet v2 models for evaluation on the heldout set and observe similar numbers.

We include the MRR numbers for other non-Duet based approaches that are available on the public leaderboard in Table \ref{tbl:results}.
As of writing this paper, BERT \citep{devlin2018bert} based approaches---\eg, \citep{nogueira2019passage}---are outperforming other approaches by a significant margin.
Among the non-BERT based approaches, a proprietary deep neural model---called IRNet---currently demonstrates the best performance on the heldout evaluation set.
This is followed, among others, by an ensemble of CKNRM \citep{dai2018convolutional} models and the single CKNRM model.
The single Duet v2 model achieves comparable MRR to the single CKNRM model on the eval set.
The ensemble of Duet v2 models, however, performs slightly worse than the ensemble of the CKNRM models on the same set.
\section{Discussion and conclusion}
\label{sec:conclusion}

In this paper, we describe several simple modifications to the original Duet model that result in significant improvements over the original architecture on the MS MARCO task.
The updated architecture---we call Duet v2---achieves comparable performance to other non-BERT based top performing approaches, as listed on the public MS MARCO leaderboard.
We note, that the Duet v2 model we evaluate contains significantly fewer learnable parameters---approximately $33$ million---compared to other top performing approaches, such as BERT based models \citep{nogueira2019passage} and single CKNRM model \citep{dai2018convolutional}---both of which contains few hundred million learnable parameters.
Comparing the models based on the exact number of learnable parameters, however, may not be meaningful as most of these parameters are due to large vocabulary size in the input embedding layers.
It is not clear how significantly the vocabulary size impacts model performance---an aspect we may want to analyse in the future.
It is worth emphasizing that compared to other top performing approaches, training the Duet v2 model takes significantly less resource and time---$1.5$ hours to train a single Duet model and to evaluate it on both dev and eval sets using a Tesla K40 GPU---which may make the model an attractive starting point for new MS MARCO participants.
The model performance on the MS MARCO task may be further improved by adding more depth and / or more careful hyperparameter tuning.

\bibliographystyle{plainnat}
\bibliography{bibtex}

\end{document}